\documentclass[a4paper,11pt]{article}
\usepackage{pos}
\usepackage[capitalize]{cleveref}
\usepackage{siunitx}

\title{Towards NNPDFpol2.0}

\author*[a,b]{Felix Hekhorn}

\affiliation[a]{University of Jyvaskyla, Department of Physics, P.O. Box 35, FI-40014 University of Jyvaskyla, Finland}

\affiliation[b]{Helsinki Institute of Physics, P.O. Box 64, FI-00014 University of Helsinki, Finland}

\emailAdd{felix.a.hekhorn@jyu.fi}

\abstract{We review the recent efforts in the NNPDF Collaboration towards a new global extraction of polarized
parton distributions functions (pPDF). Polarized PDFs are highly relevant for the interpretation of current and
future polarized high-energy experiments, including the upcoming Electron-Ion Collider (EIC).
We present a recent study of the role played by heavy quark effects in polarized DIS, where we apply the FONLL
general-mass variable-flavour-number scheme for the first time in a polarized setup and demonstrate the significant
impact of charm mass corrections, specifically for the polarized gluon distribution. We show preliminary results (DIS-only)
of this new pPDF release, NNPDFpol2.0, based on the NNPDF4.0 fitting machinery and the associated new theory prediction pipeline.}

\FullConference{31st International Workshop on Deep Inelastic Scattering (DIS2024)\\
 8–12 April 2024\\
Grenoble, France\\}

\begin{document}
\maketitle

\section{Introduction}
With the rising certainty that an Electron Ion Collider (EIC)~\cite{AbdulKhalek:2021gbh,Anderle:2021wcy} will eventually be built,
we have the chance to unravel more of the internal structure of hadrons and in particular the proton.
As an important goal we may be able to shed more light onto the resolution of the proton spin puzzle, by measuring
a significant number of longitudinally polarized cross-sections and structure functions with unprecedented precision.

The expected experimental uncertainty will allow us to constrain the polarized parton distribution functions (pPDF)
in a much broader kinematic range than the region which is accessible with current data. This is particularly relevant for the
polarized gluon distribution $\Delta g$ which, despite the direct connection to the proton spin sum rule, remains
poorly constraint.

To match the experimental precision on the theory side pPDF collaborations are challenged to improve their
pPDF determination frameworks. Here, we present the current efforts inside the NNPDF collaboration towards
a new extraction of pPDF, dubbed NNPDFpol2.0, eventually at next-to-next-to-leading order (NNLO) accuracy.

The rest of the paper proceeds as follows. In \cref{sec:hq} we review our recent paper~\cite{Hekhorn:2024tqm}
where we explore heavy quark mass effects in polarized deep-inelastic scattering (pDIS) at NNLO using for the first time a
general mass-variable flavor number scheme (GM-VFNS). In \cref{sec:pdf} we highlight the current, preliminary
status of our new extraction based on the currently available data in pDIS.

\section{Heavy quarks in polarized deep-inelastic scattering at the electron-ion collider}
\label{sec:hq}

The precise measurement of the DIS structure functions at HERA~\cite{H1:2018flt} established the necessity for a GM-VFNS,
such as the FONLL prescription~\cite{Forte:2010ta}, for unpolarized PDF extractions and we expect a similar scenario
for the polarized case with the EIC.
The basic idea of the FONLL prescription can be summarized by the formula for computing a structure function $g$
\begin{equation}
  g^{\rm FONLL} = g^{\rm FO} + g^{\rm RES} - g^{\rm sub}
\end{equation}
where $g^{\rm FO}$ is the fully massive calculation, which retains all heavy quark effects
but only contains a finite amount of collinear logarithms, $g^{\rm RES}$ is the fully massless
calculation, which contains no power-like heavy quark effects but collects the full tower of collinear resummation,
and $g^{\rm sub}$ accounts for the double-counting between the two calculations.
While the FONLL prescription~\cite{Forte:2010ta} was formulated first for unpolarized PDFs the prescription
can equally well be applied to the polarized case as it is indeed spin independent.

In fact, the only remaining step is to provide an actual implementation to compute structure functions
in the various flavor number schemes (FNS) that serve as ingredients to the FONLL prescription.
Using the \texttt{EKO}~\cite{Candido:2022tld} and \texttt{yadism}~\cite{Candido:2024rkr} libraries we are
able to do so at NNLO accuracy and, moreover, since both of them are integrated into the \texttt{pineline}
framework~\cite{Barontini:2023vmr} we are able to provide the FONLL predictions at basically no additional cost.

\begin{figure}
  \includegraphics[width=.5\textwidth]{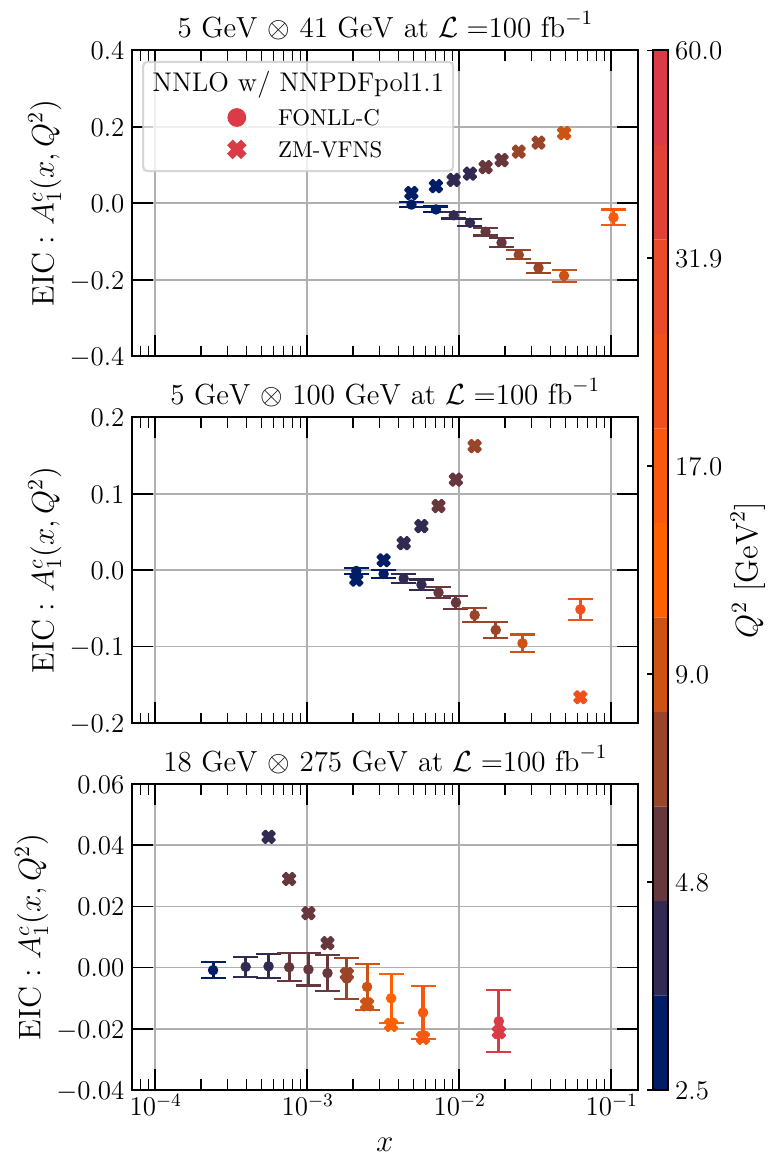}%
  \includegraphics[width=.5\textwidth]{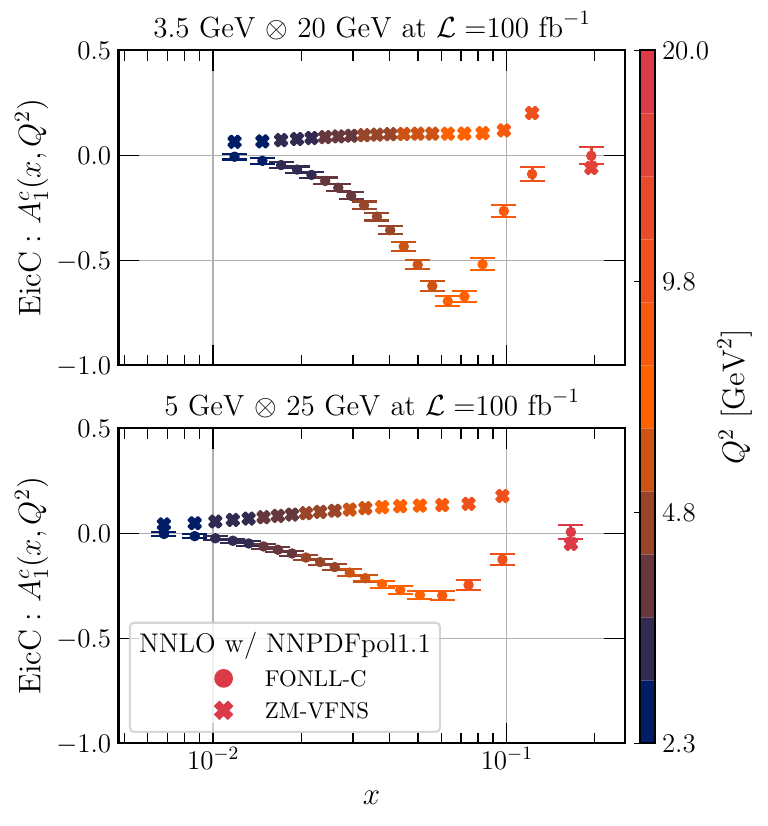}
  \caption{Theory predictions for the single-spin charm asymmetry $A_1^c$ using ZM-VFNS (crosses) or FONLL (dots) for several
  proposed kinematics at the EIC~\cite{AbdulKhalek:2021gbh} (left) or EicC~\cite{Anderle:2021wcy} (right). The indicated error bar
  corresponds to the projected experimental uncertainty~\cite{Anderle:2021hpa,Anderle:2023uvi}.}
  \label{fig:A1c}
\end{figure}

Next, we turn to the phenomenological impact by studying the single-spin charm asymmetry $A_1^c$ given by
\begin{equation}
  A_1^c(x,Q^2) = \frac{g_1^c(x,Q^2)}{F_1^c(x,Q^2)}
\end{equation}
using the (polarized) structure function $F_1^c$ (and $g_1^c$).
In \cref{fig:A1c} we compare theory predictions for $A_1^c$ using NNPDFpol1.1~\cite{Nocera:2014gqa} for several
proposed kinematics at the EIC~\cite{AbdulKhalek:2021gbh} and EicC~\cite{Anderle:2021wcy}.
We show predictions using the FONLL scheme, as proposed in Ref.~\cite{Hekhorn:2024tqm}, and the zero mass-variable
flavor number scheme (ZM-VFNS), which neglects all heavy quark mass effects and which was used in all
pPDF extractions so far~\cite{Nocera:2014gqa,deFlorian:2014yva,Ethier:2017zbq,Bertone:2024taw}.
Comparing with the projected experimental uncertainties~\cite{Anderle:2021hpa,Anderle:2023uvi} we conclude that we
will be able to resolve heavy quark mass effects at the EIC.

\section{Towards NNPDFpol2.0}
\label{sec:pdf}
Following the recent progress inside the NNPDF collaboration on methodological matters, which culminated in the NNPDF4.0
release~\cite{NNPDF:2021njg,NNPDF:2021uiq}, and the newly established \texttt{pineline} framework~\cite{Barontini:2023vmr} for theory predictions we are working
toward a new pPDF extraction, dubbed \text{NNPDFpol2.0}. Based on the work in Ref.~\cite{Hekhorn:2024tqm} we show preliminary
results based on the currently available pDIS data~\cite{EuropeanMuon:1989yki,SpinMuon:1999udj,E142:1996thl,E143:1998hbs,E154:1997xfa,E155:2000qdr,
JeffersonLabHallA:2016neg,Kramer:2002tt,JeffersonLabHallA:2004tea,CLAS:2014qtg,COMPASS:2006mhr,COMPASS:2010wkz,HERMES:1997hjr,HERMES:2006jyl}.
As we only have cross-section using neutral current (NC) interactions between the electron and
the proton available, we have limited sensitivity to the flavor decomposition and, thus, we choose to parametrize
\begin{equation}
  \Delta g,\quad \Delta \Sigma, \quad \Delta T_3 = \Delta u^+ - \Delta d^+, \quad \Delta T_8 = \Delta u^+ + \Delta d^+ - 2\Delta s^+
\end{equation}
at our parametrization scale $Q_0 = \SI{1}{\GeV}$.

In addition to the experimental constraints we also impose in addition three theory constraints:
\begin{enumerate}
  \item the absolute value of the pPDF have to be bound by their unpolarized counterpart (here NNPDF4.0~\cite{NNPDF:2021njg})
    at $Q^2 = \SI{5}{\GeV^2}$~\cite{Candido:2023ujx,Hekhorn:2024foj}
    \begin{equation}
      \left|\Delta f(Q^2 = \SI{5}{\GeV^2})\right| < f(Q^2 = \SI{5}{\GeV^2})
    \end{equation}
  \item the first moment of $\Delta T_3$ and $\Delta T_8$ is constrained by baryon decays~\cite{ParticleDataGroup:2022pth}
    \begin{equation}
      \int\limits_0^1\!dx \Delta T_3(x,Q^2) = a_3, \quad \int\limits_0^1\!dx \Delta T_8(x,Q^2) = a_8
    \end{equation}
  \item the first moment of $\Delta g$ and $\Delta \Sigma$ is finite
  \begin{equation}
    \int\limits_0^1\!dx \Delta g(x,Q^2 = \SI{1}{\GeV^2}) < \infty, \quad \int\limits_0^1\!dx \Delta \Sigma(x,Q^2 = \SI{1}{\GeV^2}) < \infty
  \end{equation}
\end{enumerate}

\begin{figure}
  \begin{center}
    \includegraphics[width=.45\textwidth]{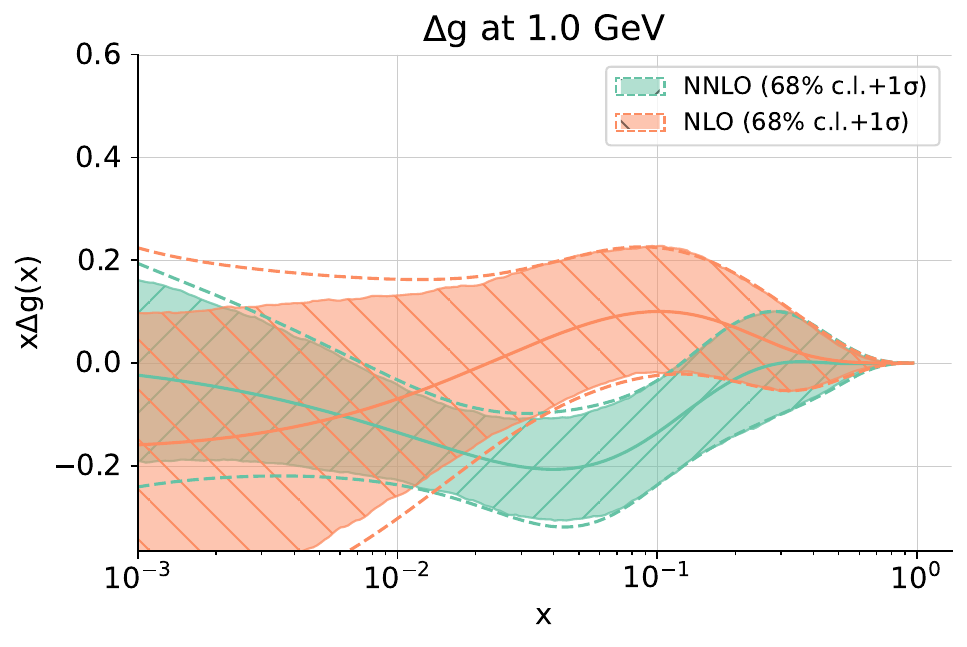}%
    \includegraphics[width=.45\textwidth]{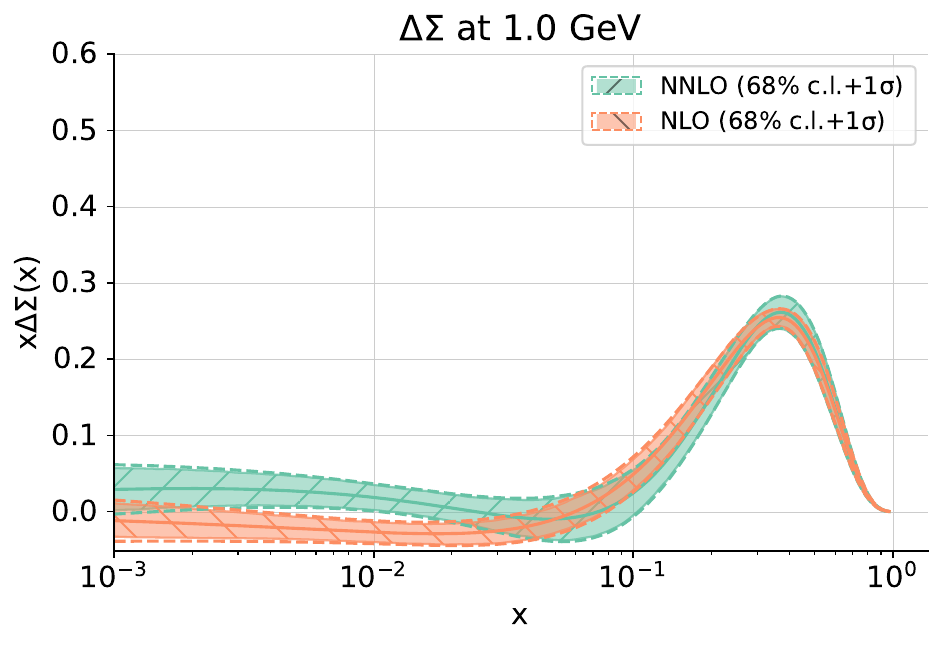}
    \includegraphics[width=.45\textwidth]{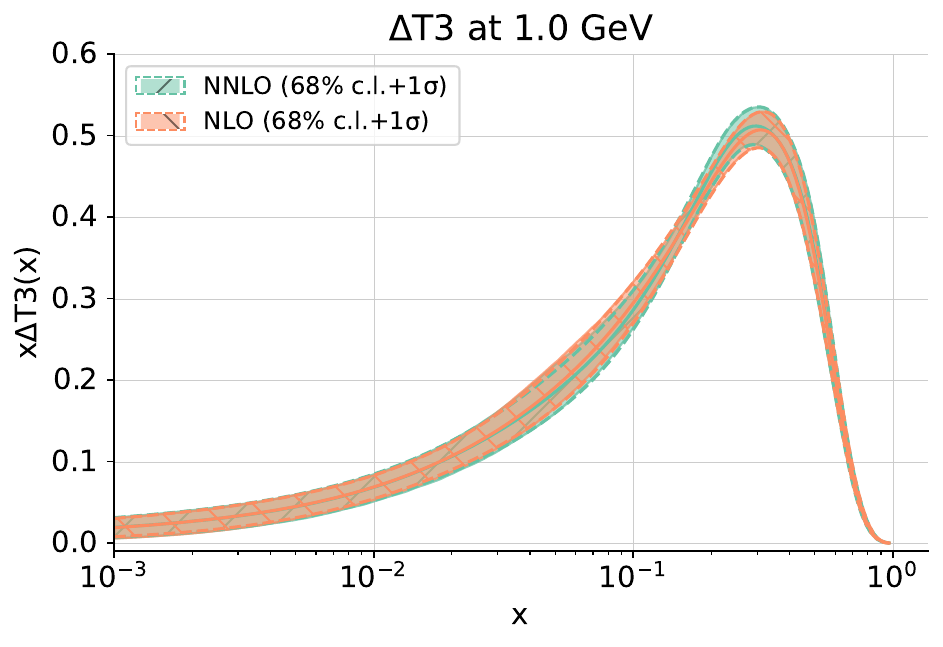}%
    \includegraphics[width=.45\textwidth]{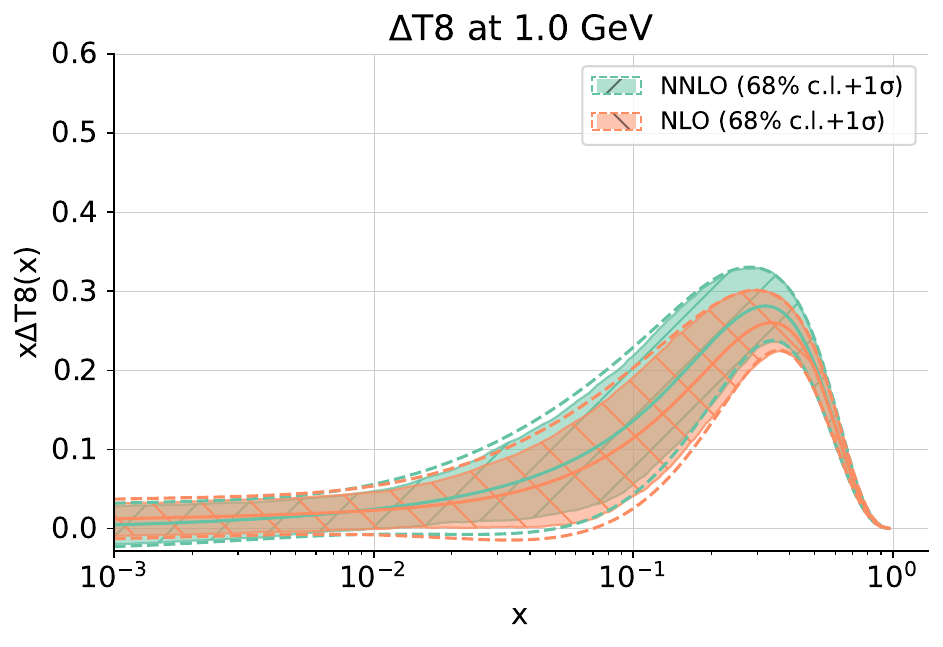}
  \end{center}
  \caption{Preliminary results for the PDFs of NNPDFpol2.0 as a function of the momentum fraction $x$ at the fitting scale $Q_0 = \SI{1}{\GeV}$.
  From left to right and top to bottom we show the polarized gluon $\Delta g$, singlet $\Delta \Sigma$,
  singlet-like triplet $\Delta T_3$ and singlet-like octet $\Delta T_8$ at NLO (orange) and NNLO (green) accuracy.}
  \label{fig:PDF}
\end{figure}

With this setup established we perform a first preliminary pPDF determination at next-to-leading (NLO) and NNLO accuracy and we show the
resulting pPDFs in \cref{fig:PDF}. We observe that the obtained results match the expected pattern: while we can indeed already constrain
the quark pPDFs to a certain degree, as they may couple directly to the exchanged boson, the gluon pPDF still remains fairly unconstrained.
Recall that the gluon only enters via higher order corrections, i.e.\ $O(\alpha_s)$ and above, or scaling violations, i.e.\ via the DGLAP equations,
into DIS cross-sections. We also note a relevant difference between the NLO and NNLO determination of $\Delta g$, which might be related
to the FONLL prescription, as there we have a higher sensitivity to the gluon distribution as we retain the fully massive calculation,
which at leading order (LO) is given by photon-gluon-fusion.

Moving forward towards a full-fledged pPDF release we are planning to include more diverse datasets, specifically from purely hadronic
collisions such as weak boson production~\cite{STAR:2018fty} or (di-)jet production~\cite{PHENIX:2010aru,STAR:2012hth,STAR:2014wox,
STAR:2016kpm,STAR:2018yxi,STAR:2019yqm,STAR:2021mfd,STAR:2021mqa} at RHIC. This will be particularly helpful to further constrain the
polarized gluon distribution $\Delta g$. We also plan to account for the uncertainty associated to the finite perturbative knowledge,
referred to as Missing Higher Order Uncertainty (MHOU), as was recently demonstrated for NNPDF4.0~\cite{NNPDF:2024dpb}.

\acknowledgments
F.~H. is supported by the Academy of Finland project 358090 and is funded as a part of the
Center of Excellence in Quark Matter of the Academy of Finland, project 346326.

\bibliographystyle{JHEP}
\bibliography{ref.bib}

\end{document}